\documentclass[sigconf]{acmart}

\setcopyright{acmlicensed}
\copyrightyear{2026}
\acmYear{2026}
\acmDOI{NA}
\acmConference[Submitted to Future of Software Engineering]{Submitted to Future of Software Engineering at ICSE 2026}{January 2026}{Rio, Brazil}
\acmISBN{NA}

\acmSubmissionID{TBC}

\begin{document}

\title{On the Abolition of the ``ICSE Paper'' and the Adoption of the ``Registered [Proposal]'' and the ``[Results] Report''}




\author{Fabio Massacci}
\authornote{Both authors contributed equally to this research.}
\affiliation{%
  \institution{Univ.\ of Trento, IT and Vrije Univ.\ Amsterdam, NL}
  \city{}
  \state{}
  \country{}
}
\orcid{1234-5678-9012}
\email{fabio.massacci@ieee.org}

\author{Winnie Bahati Mbaka}
\authornotemark[1]
\affiliation{%
  \institution{Vrije Univ. Amsterdam, NL}
  \city{}
  \state{}
  \country{}
}
\orcid{1234-5678-9012}
\email{w.mbaka@vu.nl}

\renewcommand{\shortauthors}{Fabio and Winnie}

\begin{abstract}
  To address the 'novelty-vicious cycle' and the `replicability crisis' of the field (both discussed in the survey) we propose abolishing the ``ICSE paper'' as we know it and replacing it with a two-tier system that also evolves the existing notion of `Registered Report'. Authors proposing a new idea, experiment, or analysis would submit a ``Registered [Proposal]'' of their idea and the proposed experimental methodology to undergo peer review. The following year, \emph{anyone} can submit (shorter) ``[Results] Reports'' on the realization of the empirical work based on the registered proposals of the previous ICSE (or FSE or ISSTA or ASE etc.). Both works should be first class citizens of the mainstream events. We argue that such a disruptive (heretical?) idea is supported and based on the responses of the community of the Future of Software Engineering pre-survey.
\end{abstract}




\received{14 January 2026}
\received[revised]{possibly}
\received[accepted]{maybe}

\def\fsequote#1#2{%
\begin{center}%
\noindent\parbox{0.9\linewidth}{$\circ$ {\itshape #2} (#1)}%
\end{center}%
}
\maketitle

\pagestyle{plain}

\section{Introduction}

There is a broad agreement that there is a replicability crisis in software engineering \cite{robles2010beyond,madeyski2017would}. This is starkly visible in the field of Machine Learning for vulnerability detection, where each year new papers show that previous results are fundamentally wrong \cite{risse2025top}. This feeling is also broadly reflected in the comments collected from the survey on the Future of Software Engineering \cite{ICSE26}. 
We select two themes to highlight the two major problems of the field that we will address in the rest of the position paper. The first theme is indeed reproducibility:
\fsequote{192758693}{We are not good at replication (despite better rigour in what we do), which is partly a publication thing (journals and conferences do not rate these). Too much effort goes into chasing `new' rather than consolidating our knowledge.}

We need, therefore, to transform our publication model in a way that still values conference submissions for novelty but makes it possible to validate the results by \emph{somebody other than the inventor}. 

The second problem is the elitist nature of the research community. Although a derogatory interpretation might be that there is a mafia:
\fsequote{192154506}{Research \ldots favor established names and groups}
\fsequote{192676762} {ICSE is very cliquey}
We want to propose a more general interpretation and, in our opinion, closer to the truth interpretation:
\fsequote{192295956}{there is still a strong fear of trying innovative ideas that might be perceived as challenging the work of well-established and renowned researchers in the field.}

Combined with the replicability crisis this generate a negative feedback loop: all-important ideas are published by the all-important groups, nobody can essentially show that the original idea was wrong (as negative results are not appreciated) \emph{unless} the challengers invent also something new, and the new idea cannot be replicated or negated as you again need to have something new\ldots As pointed out also in the Discord discussion:
\fsequote{Discord}{One notable ``problematic area''  is that we focus on "novelty" so much that we create many tool prototypes that are unusable in practice - ones thar require painful setup and work mainly on ``toy examples''.}

 As a conclusion research results make very limited impact:
\fsequote{196067582}{I think this makes SE research body of knowledge very shallow and with little relevancy to the industrial practice.}

This is unsurprising: if researchers themselves cannot replicate new solutions, how should practitioners adopt them? 
To propose a solution to this impasse we need a bold move and our proposal is to abolish the traditional ``ICSE paper'' and replace it with two different type of papers (the ``Registered Proposal'', which partly exists in the form of a ``Registered Report'') and the ``Results Report'' which does not exist yet and we will introduce here.

\section{[Taking stock] - Problems and Trade-Offs}

We review here some of the results of the survey on the Future of Software Engineering \cite{ICSE26}. While an interesting theme is surely managing reviews and scaling down both numbers and stress\footnote{The word 'review*' occurred 118 times just in the question about if you could change one thing what you could do with ``write 1, review 3'' proposals.} we believe two interesting themes emerged;
\begin{itemize}
    \item Novelty vs Rigor
    \item Replication vs Novelty
\end{itemize}
In the first theme, a perceived trade-off between novelty and the rigor of empirical validation is implied. On the one hand, Software Engineering communities view novelty as either the use of new datasets, the development of a new tool, or a new technique. On the other hand, validation is evaluated after the fact and is conditioned on the strength of the published results. 

\fsequote{196022336}{the empirical evaluation seems to be more important than the approaches themselves.}
\fsequote{193757741}{The excessive focus on rigor and big samples seems to have made it harder to connect with real practitioners}
\fsequote{195811076}{Reviewers should look for new ideas in papers, now only for the most thorough evaluations.}
\fsequote{195844264}{Be more open about new ideas and mitigate requirements on tooooo conventional and rigorous points}
\fsequote{196077577}{reviewers \ldots hold firmly to their own assumptions or preferred methodologies. This can lead to biased evaluations, where papers proposing new directions or challenging established norms face unnecessary resistance.} 
In a nutshell, we can summarize the problem as
\fsequote{196196304}{The review and publication culture is becoming overly focused on incremental but sound engineering advances, at the expense of novel/disruptive ideas [\ldots] should be more open to contributions not yet fully validated in terms of external validity, but with great novelty potential} 

In the second theme, a perceived tension between the Software Engineering community's commitment to rigor through replication and the structural process of peer reviews governing publications is implied. 
Overall, a much larger number of respondents of the survey \cite{ICSE26} pointed out that replication studies remain systemically marginalized due to the focus being on novelty. \fsequote{196067582 - our emphasis}{we are reluctant to accept papers that replicate previous study, or just provide evidence in the areas that were already studied \textbf{(sometimes having single studies only)} as we believe these are not new results.}
\fsequote{196247974}{A reviewing practice\ldots accepting novelty also comes with replications.}
\fsequote{196290588}{be more open for paper submissions that may be just about specific case studies, or have negative results, or just replicate previous studies under some different settings.}
\fsequote{196144420}{too much focus on \ldots moving to the next hot topic, instead of ensuring reproducibility}

Of course there are increasing efforts towards replication and negative results (e.g. ICPC'26 session, the RENE session at some ICSEs, the MSR and ESEM registered reports) but they are \emph{not} papers in the main track and sometimes \emph{not even in the proceedings}. They are always children of a lesser God. 
\fsequote{192758693}{Not sure how they could be handled by conferences, but surely we can devise some way of giving researchers credit for such studies.}

Since there are no established incentives to promote independent replication studies, there is a shift in how trust is established, via reputation, whereby papers from well-known authors and research groups rarely undergo subsequent criticism. 
\fsequote{192295956}{we often become trapped in this cycle unless there is collaboration with a well-known researcher, almost as a form of authorization or blessing}

In addition to the responses to the survey \cite{ICSE26}, we also reviewed the discussion held on the community Discord channel. While the majority of the discussion focused on other topics, a small number of comments reiterated related concerns towards novelty vs full replication (see quote in the introduction).

\section{[Planning for change] The Proposal}
So we start from these two key hints from the survey participants:
\fsequote{196205708}{A community where high-quality, durable artefacts and real-world impact are valued as much as novelty}	
\fsequote{196069944}{I would change the review process to heavily weight and reward the replication and extension of prior foundational work, hoping this would establish more reliable results and make SE research cumulative rather than trend-driven.}

We propose to 
\begin{enumerate}
\item Abolish the ``ICSE technical paper'' as we know in which \emph{both} idea, methodology, \emph{and} corresponding validation are performed by the \emph{same} people.
\item Introduce a ``Registered [Proposal]'' which should focus on the \emph{novelty} proposing new research questions, hypotheses (where applicable), and a detailed methodological and analysis plan.
\item  Introduce a ``[Results] Report'' which should focus on the \emph{rigor} where \emph{anyone} can submit a validation experiment of a previously published ICSE Registered Proposal.
\end{enumerate}
Our view both ``Registered [Proposal]'' and ``[Results] Report'' of the previous year's proposals are officially part of the main track (also because there are no more ICSE papers!). The latter report would be shorter papers, which would be complemented by the artifacts. They can be shorter as they don't need to discuss the methodology. This could be just a short half-page pointer to the proposal.

A key aspect here is the submission by \emph{anyone}. This is very different from the current model of the Registered Reports (where only the submitters of the original report are granted access to the journal submission) and we will return to discuss it.

Reports could also be organized at the conferences as a session around the proposal they validated. We expect that the result will be confirmatory in some cases (most likely when done by the proposers) but we are willing to bet money that in most cases expectations will not be confirmed (when done by somebody else). Only in few cases the results will hold and those will be the papers industry should be keen to see. Those should also be the ones to be awarded.

We say ``ICSE paper'' as a placeholder for a technical paper of the main research track at a software engineering conference. It could be interpreted as the ASE, FSE, ISSTA, etc. \ldots, paper. Since we need to start somewhere, let's start from the ICSE papers.

\subsection{Addressing Novelty}

Existing Registered Reports are split into confirmatory and exploratory studies, but there is an expectation of the results:
\begin{quote}
\texttt{Exploratory studies in software engineering often cannot be adequately assessed until after the study has been completed and the findings are elaborated and discussed in a full paper. [\ldots] 
It is difficult to offer IPA, as we do not know whether it is any better than a traditional approach based on e.g., decision trees.} (Excerpt from MSR/ESEM call for Registered Report)
\end{quote}
We argue that this expectation is actually misplaced. The whole purpose of our proposed `Registered Proposal' is to actually propose \emph{novel and interesting ideas}. They should be methodologically well-grounded in past research, and there should be some theoretical or preliminary empirical justifications for why the new idea should work. For example, which gap should close that past experiments and negative results (that are now available) have shown not to work. They should also have methodological rigor in the \emph{proposed} evaluation. They might even have code to process the stats of the results if they are available in a suitable template. 
\fsequote{192295956}{our role as scientists is precisely to question why things are the way they are and whether they still make sense.\ldots 
I would like to see a community where emerging researchers feel encouraged and supported to challenge established ideas and to innovate freely, without fear of invalidation or gatekeeping.}

For the validation part the proposal should give \emph{sufficiently specific and yet sufficiently broad} criteria for the artefacts to be the object of rigorous experimental validation.

To check whether they will work or not (as opposed to could work) should be the results of the `Results Report' Papers.

\subsection{Addressing Replicability}

Once the proposal is accepted and presented, the race starts and is open to all. Everybody could try to show that it works on any set of  artifacts that meets the criteria.  Submitters will have to show that they follow the protocol but most importantly that they have valid artifacts. Remember, we are addressing the problem of replicability. 

\fsequote{196020447}{I think we should invest heavily in reviewing software artifacts and reject papers if the artifacts do not support the storyline. At least for papers dealing with software systems, artifacts often allow the rerun experiments. If the artifact is not evaluated to be functional, the paper must not be accepted, IMHO.}
\fsequote{196016981}{Rigorous and systematic requirement for replication packages in a standard, easily executable format.}    	
We are uncertain whether there should be \emph{in principle or (continuity) acceptance} as in journal submission procedure for Registered Report for the same authors of the Report. This bonus point make sense because RR are \emph{not} in the official proceedings. If there was not at least the promise of some official publication venue nobody would ever submit a Registered Report. In our set-up the authors of the Registered Proposal would have a head-start over the others, but they should also be held to the same standards as the other. We believe that in our set-up, this would not be necessary. One of the main uncertainties is the question of reviewers asking to follow their pet methodologies or artifacts:
\fsequote{196077577}{reviewers \ldots hold firmly to their own assumptions or preferred methodologies.}
In the new model, that risk would be gone. Reviewers asking to change protocols or criteria for experimental targets because they disagree with the ``Registered Proposal'' should be simply kicked off by the chair. They should have reviewed the proposal! 

By allowing the same registered report to be executed by multiple teams, replication is \emph{injected} directly into the publication pipeline rather than being approached as an afterthought. In such cases, negative or conflicting results can be viewed as informative and present a way for robustness to be measurable rather than assumed.

\fsequote{192249436}{I would like to see an avenue to express negative work in a respected way. 'I tried x and it just didn't work out well' \dots if the effort was well designed, the evaluation sound and the conclusion just happened to be 'nope'}
\fsequote{196247974}{Dissemination of negative results. Replication towards consolidation of ideas and solutions}

Replication by anyone would also address some of the problem of replicability raised by respondents
\fsequote{196196304}{I find many cases in which reproducibility is only 'apparent', with researchers even removing files from their Github repos after publication, making their work inaccessible or unusable. I don't have a solution, but this is indeed a barrier to openness.}

\section{Conclusions?}
In this paper we have tried to address the challenges currently faced by the software engineering community to balance novelty with reproducibility to keep industrial relevance. 

Such analysis is always based on interpretations. Indeed we can claim that the community has made great steps forward towards replicability \fsequote{196077577}{One aspect that works well is the community’s increasing emphasis on sharing datasets, code, and artifacts. This openness supports reproducibility, allows others to build on prior work more easily, and helps establish shared benchmarks}
Yet, this optimistic view is challenged by many other respondents
\fsequote{196173071}{One aspect that does not work well \ldots is the limited reproducibility of many published studies. Complex toolchains, unavailable datasets, and insufficient documentation often make it difficult for other researchers to replicate results.}

We do believe the community has made significant progress but still significant work remains. Our proposal is step towards giving replicability its due as a first class citizen and separating such due from novelty.

Of course our proposal may fail to deliver its goals. Yet, we are confident it would at least solve the long standing open problem of the `post-doc lost hard disk': when you contact the all-important professor of the all-awarded team to obtain a copy of the all-showing data and code, when you get the answers, it turns out that the post-doc has lost the hard disk. By having different teams submit different replications, it is exponentially unlikely that they will all lose the hard disk. The surviving hard disks might even show that the all-important proposal did not work after all.

If our proposal succeeds to deliver, we would have at least two independent groups submitting a successful validation and we would have reached the IETF rule for a RFC to be adopted as a successful technology ``two independent interoperable implementations'. Industry adoption would naturally follow.

\section*{Acknowledgment}
We would like to thank the anonymous participants in the Future of Software Engineering Survey. Your suggestions and your wordings were invaluable to give a coherent shape to our vague feelings and to give form to this idea. If you recognize your quotation and think that is not what you intended, let us know. 

This work has been partly supported by the European Union Horizon Europe Program under grant 101120393 (Sec4AI4Sec), by the Italian Ministry of University and Research (MUR), under the P.N.R.R. – NextGenerationEU grant n.\ PE00000014 CUP E63C24000590001 (SERICS subproject COVERT) and the NWO under grant KIC1.VE01.20.004 (HEWSTI).

The authors declare that they have no conflict of interest.
\subsection*{CRediT statements}
	\emph{Conceptualization:} FM, WM ; 
	\emph{Methodology:} FM, WM ; 	
	\emph{Software:} NA ; 
	\emph{Validation:} FM, WM ;	
    \emph{Formal analysis:} FM, WM ;	
    \emph{Investigation:} FM, WM ;	
    \emph{Resources:}NA; 
    \emph{Data Curation:} FM, WM ; 	
    \emph{Writing - Original Draft:} FM, WM ;
    \emph{Writing - Review \& Editing:} FM, WM ; 
    \emph{Visualization:} NA; 
    \emph{Supervision:}  FM ; 
    \emph{Project administration:}  FM ; 
    \emph{Funding acquisition:}  FM ; 

\bibliographystyle{ACM-Reference-Format}
\bibliography{sample-base}

\appendix

\section{Raw Quotes}

For reproducibility, we list here all raw quotes from the Survey on the Future of Software Engineering from \cite{ICSE26} that included the keywords novel*, new, rigor, reproduc*, replic* that are at the basis of our analysis. There are 53 statements by different particpants, of which 11 made two different statements with the relevant keywords. The statement are sorted in order of survey ID.

We included also two relevant citations on the corresponding discussions on the Discord channel. 

\clearpage 

\onecolumn
\fsequote{192109802}{Increasing expectations around data sharing and replication packages.}

\fsequote{192134556}{Peer review suffers from poor alignment with collective progress and real world practice. It has for as long as I have participated but the problem has worsened recently. Reviewers latch on to minor imperfections to reject more creative/novel papers, which significantly depressed my students’ (I’m also faculty) interest in academia and has steered most AI for code research to other venues, to give an example. I frequently have to argue at length to accept interesting papers with minor flaws.}

\fsequote{192143242}{I think the conference review committees, especially ICSE and FSE, are reluctant to welcome new reviewers. I have participated as a shadow reviewer in multiple MSR conferences during my PhD and have served as a reviewer for a journal and several workshops. However, even though I have applied and contacted the chairs of these committees about opportunities, I have never been given a chance. It often feels like the same committee members appear repeatedly, and when there is a newcomer, it is usually someone the chair already knows well. As a result, there seems to be little room for new reviewers to join.}

\fsequote{192250097}{Reviewing workload and its schema: the model is not balanced and does not reward reviewers. The golden access policy implemented by most publishers is unfair and fails to guarantee access for small/poor departments. The reviewing model of most conferences mimics journals with rebuttals and major revisions, leading to publishing delays and wasting reviewers' efforts. I love the open science policy with open-source replication packages.}

\fsequote{192295956}{I feel that, especially in my country, there is still a strong fear of trying innovative ideas that might be perceived as challenging the work of well-established and renowned researchers in the field. However, I believe that our role as scientists is precisely to question why things are the way they are and whether they still make sense. As time passes, society changes, and so do our needs. My impression is that, as students and early-career researchers without a famous name, we often become trapped in this cycle unless there is collaboration with a well-known researcher, almost as a form of authorization or blessing to propose something disruptive. If I could make one change, it would be to foster a culture that truly values originality and critical inquiry, regardless of seniority or reputation. I would like to see a community where emerging researchers feel encouraged and supported to challenge established ideas and to innovate freely, without fear of invalidation or gatekeeping.}

\fsequote{192311611}{Their unwillingness to have reasonable exceptions. They expect short papers to have every detail like full paper. Then, they expect the full paper should have more detail like journals. And when provided replication packages, they dont even open it.}

\fsequote{192311611}{Too much focus on novelty and not actual substances. It feels like the community think if a paper is accept and the novelty of the topic is gone and no need to explore it further. The reviews with the advert of ChatGPT went batshit recent days in this regard. Secondly, the failure of adopting new technology, new adaptations. For examples, the community would do review with ChatGPT but won't disclose but they won't accept HitL/LLM-as-a-Judge annotation mechanism. In order to convince the community, it seems like we need to do NLP research as well, as if doing two research in one go. I would not be surprise if people try to find venues in other domain rather than SE research community, given how difficult reviewers make it.}

\fsequote{192587927}{The ICSE community, while vibrant and welcoming, is not as open to new ideas as one might expect when compared to conferences in other areas of computer science. This relative insularity has consequences: when engaging with researchers outside the software engineering community, we are often perceived as less relevant, less impactful, and overly focused on industrial applications. For example, the influence of the AI community on our work is far greater than our influence on theirs. Many of our innovations remain largely unknown outside software engineering, while we frequently adopt and adapt breakthroughs from other areas—such as supervised and unsupervised learning, NLP strategies, and, more recently, large language models. This raises a fundamental question: how can we position the ICSE community so that other research fields see us as influential and forward-looking, rather than merely consumers of external advances? We often express concerns about the relatively small size of subcommittees and program committees. However, we are not always proactive in inviting reviewers from outside our immediate field. Given the rapid rise of AI, for instance, there is a growing need to involve reviewers with expertise in AI and related areas. Expanding our reviewer base is just one step toward fostering greater openness and cross-pollination of ideas. Despite these challenges, our community is notably friendly, supportive, and full of untapped potential. By consciously improving the values we promote internally—such as inclusivity, recognition of innovation, and collaboration—we can significantly enhance our external visibility and competitiveness. Many other research communities are widely recognized for specific achievements, with LLMs being a clear example. In contrast, we have a wealth of success stories, both past and present, that are often under-communicated. Highlighting the novel approaches, automation tools, and influential software projects developed within our community would strengthen our profile.
A lack of strong panels and high-visibility discussions further diminishes our ability to broadcast these achievements. Observing other conferences, it is clear they are highly effective at promoting their contributions and shaping their public image—something from which we could learn. Another area for improvement is participation in standardization efforts. Currently, our presence is minimal. With the growing impact of AI, there is a pressing need for our community to actively contribute to standardization discussions. Engaging in these efforts would not only elevate our relevance but also create valuable opportunities for joint exploration with industry and other research fields. More broadly, attending an SE conference should provide attendees with a clear understanding of: * What our community has achieved and how we can build upon these successes. * Where knowledge gaps and open challenges remain. * The impact goals we aim for in the future—not just reflecting on past achievements, but charting a vision for the next era of software engineering research. By addressing these points—embracing openness, celebrating our achievements, engaging in standardization, and clarifying our future impact—we can ensure that ICSE is recognized as a truly influential and forward-thinking community, both within computer science and beyond.}

\fsequote{192604904}{These is perhaps more friction to publishing some research than necessary; the field needed much more statistical rigor and standards (non-toy examples) but has perhaps become too insistent on certain shibboleths, sometimes improperly applied.}
\fsequote{192758693}{Greater emphasis on the importance of replication. I would like to see journals in particular have sections for replicated studies because this is how science progresses. Not sure how they could be handled by conferences, but surely we can devise some way of giving researchers credit for such studies.}
\fsequote{192758693}{We are not good at replication (despite better rigour in what we do), which is partly a publication thing (journals and conferences do not rate these). Too much effort goes into chasing 'new' rather than consolidating our knowledge.}
\fsequote{193668599}{The community is getting harser in reviews and less open to new ideas; attacks on diversity and inclusion in the US} 
\fsequote{193757741}{Tech transfer and building partnerships between industry and research. While it happens sometimes (particularly in the context of industrial research labs) and there is an important role for fundamental research not intended to be immediately practical, there's not enough emphasis on relevance to real practice for SE research intended to be practical (e.g., studies of practice, developer focused tools). This seems largely due to most in industry, even and especially those designing new developer tools or offering advice on practice, having little interest in the work SE researchers do.}
\fsequote{193757741}{The excessive focus on rigor and big samples seems to have made it harder to connect with real practitioners through empirical studies. When developing new tools and techniques, industry best practice (e.g., the Y Combinator approach to startups or internal dev tools in FAANG) is to find pressing top of mind problems, build simple solutions and ship fast, and iterate to ensure solution really addresses problem. While SE research sometimes does all of these steps, through studies of practice, building solutions, and running lab studies, in reality there's often not sufficient fit to pressing problems, unrealistic assumptions, and limited context in evaluations that lead most in industry to ignore most SE research (with many important exceptions, largely done in partnership with industry research labs).}
\fsequote{194032046}{Peer review loads are getting too high and seem to have more variation. Openness to new ideas seems relatively low.}      
\fsequote{194220564}{Rare toxic encounters/behaviours, unconscious biases, aversion to brand new ideas, not enough engagement in reflective practices (unless led by established leaders, like this one), inadvertently leaving out some groups (eg Africa, global south) from key communities activities, opportunities, and events}
\fsequote{195394675}{Funding is very 'one-sided' in some countries, a lot of 'know-how' is missing. Many AI-generated papers (and reviews), and often poor data (questionable reproducability) A lot of 'delta improvement', some are still stuck in old methods compared to industry, lack of novelty, high repeatability in what is published (same thing - new words/context). Also that people are NOT presenting live at conferences - (the uses of proxy people who cannot answer questions is very poor)... }
\fsequote{195756655}{Way out of hand from industry. Codex/copilot can eliminate the novelty of more than half of the paper’s idea}
\fsequote{195775809}{Many recent SE studies rely on LLMs, but their high computational cost makes it difficult for researchers without industry collaboration to run large-scale experiments. As a result, companies with extensive GPU resources gain disproportionate influence over the direction and reproducibility of SE research.}
\fsequote{195811076}{Change submission guidelines to foster more discussions, at workshops and at conferences. Reviewers should look for new ideas in papers, now only for the most thorough evaluations. Also make it clear that previous discussion of some research at some workshop is encouraged, and may not hurt a subsequent conference publication. The outcome would hopefully be that workshops are better attended again, and that the conferences would be more interesting, because their program would focus on 'good ideas'.}
\fsequote{195811076}{There is a big influx of AI-related papers. I particularly dislike studies of the form 'we tried something with an LLM, here it is'. While some such papers might be sensible, I feel that we have accepted too many for them into conference proceedings. This is because many use very questionable, highly biased experimental methods. This has the threat of creating an 'AI bubble' in our own community. Also I feel that generally we are all reviewing too many papers these days. There are many deadlines, with many resubmission. Part of it has to do with the fact that we have raised the standard for acceptable papers. When I did my PhD 15-20 years ago, it was often good enough for a paper to have a great idea, even if the experiments might have been preliminary. At our major conferences today this would no longer fly. I resent that. Conferences should be all about ideas, and exchanging ideas. Journals are for publishing 'your final word' on a topic, not conferences. I thus think that we should generally lower the standards and look more for good ideas in papers. Just when it comes to AI we could use a bit more rigor. Similarly, I have repeatedly now seen PhD students struggle with publishing conference papers that extend their own work, which was previously published as an idea at a workshop. 15 years ago this was the norm: you would present and discuss a research idea at a workshop then go back home, build and evaluate the thing, then publish at a conference. These days, reviewers repeatedly reject such submissions 'because the work has already appeared at this and that workshop'. I think this is insane. That way we are discouraging early discussions about research, and we are seriously hurting workshops, which allow for such discussions. I think in ACM/IEEE we should have a written rule that workshop publications may not hurt subsequent conference publications.}
\fsequote{195844264}{Attitude to new techniques should develop quickly to catch up with latest performance of advanced techs.}
\fsequote{195844264}{Be more open about new ideas and mitigate requirements on tooooo conventional and rigorous points, e.g., not accepting contest tasks' code as subjects and forcing the evaluation with dirty open source project code.}
\fsequote{195853386}{Most of the published SE research is irrelevant for practitioners, and this got worse over the past years. “We improved x\% over a random and toy baseline” is not something practitioners care about. Also, a lot of recent SE research is hype-driven, and I miss the scientific rigor and distance required for an A* conference such as ICSE. “We have thrown an LLM at a problem, and it mostly worked” is not research. Also, we're still organizing conferences in countries that exclude people from certain regions to attend, while we talk a lot about diversity and inclusion at these conferences. The option to, e.g., present an ICSE paper at the following FSE or ASE is a first step toward mitigating this (event though it's just a workaround). Our conferences are still very much focused on North America and Europe, while a lot of innovation is happening in Asia at the moment.}
\fsequote{195853386}{No paper acceptance without artifact evaluation (or a review of a statement why the artifact cannot be published). Most of the SE research being published is neither reproducible nor replicable.}
\fsequote{195892827}{Discussing potential novel solutions with fellow researchers, including PhD candidates. I engage with that too rarely (maybe one hour per week)}
\fsequote{195938395}{overuse of LLM , expectations around experiments (dataset size) that favour larger groups, and reporting incremental progress over interesting new ideas -- a lot of papers have become very 'boring' (our version of AI slop)}
\fsequote{195984732}{1) Overemphasis on novelty at the expense of rigor. Novelty and impact are explicit evaluation criteria in every call for papers for our major conferences, and perhaps we should question that again. 2) Limited diversity in methods (which ultimately leads to a lack of expertise among reviewers and thus to a lot of frustration). 3) Reviewing fatigue: too many submissions handled by too few people.}
\fsequote{196016981}{Rigorous and systematic requirement for replication packages in a standard, easily executable format. The outcome would be less blah blah and more instantly confirmable and improvable research} 
\fsequote{196020447}{I think we should invest heavily in reviewing software artifacts and reject papers if the artifacts do not support the storyline. At least for papers dealing with software systems, artifacts often allow the rerun experiments. If the artifact is not evaluated to be functional, the paper must not be accepted, IMHO. At least this should be the policy for software related papers. I understand fully that artifacts based on surveys, etc. have to be treated differently.}
\fsequote{196022336}{1) There are some 'famous old white men' in the community that disencourage others. Bully their PhDs, tell younger colleagues what they do has no value or is just wrong. And others still support them as they are the 'famous' people winning awards. This requires younger people to hide a bit in the shadows to not being attacked. 2) The “core” community seems to be a quite closed club. For newer people it’s not easy to get in or in touch. 3) The topics we are working on are extremely far from each other. I’d say SE is not “one community” any more. It’s different communities coming together sometimes as ICSE, ASE or FSE. But our communities are in the branches of SE topics. This is not well reflected in our main conferences. There shall be dedicated tracks and spaces for all branches. 4) ICSE is not an extremely nice place to be. It’s too many people, crowded, too many topics one is not interested in, too much marketing in articles (also their titles), the empirical evaluation seems to be more important than the approaches themselves. 5) Some topics get way too important and ICSE publications are basically running behind trends. It would be nice to have non-LLM tracks with papers that have no GenAI in them. 6) Conference costs get out of proportion. ICSE is extremely costly (both, conference fees and hotels) and this is not inclusive any more. With normal budgets, people could to it maybe once every 5-6 years. So it will never be a 'community forming' event. Maybe one should think about having conferences every year rather on different continents to reduce travel costs \& improve sustainability and then some cross-cutting conferences with all together every 5 years}
\fsequote{196044669}{What brings me the greatest joy in my work is uncovering novel findings that not only advance our understanding of software engineering but also have real-world impact. I take pride in ensuring my research is statistically robust and generalizable, so it can be trusted and applied beyond just theory. As an early-career researcher, there’s nothing more rewarding than seeing my work bridge a knowledge gap—it makes all the effort feel meaningful. I also find deep satisfaction in serving on program committees or as a journal reviewer. It gives me a chance to contribute to improving the review process itself, pushing for constructive, timely feedback—the kind of change I’d love to see more of in the community. While I engage with these aspects regularly, the joy is especially strong when a study comes together or when I can help shape a fairer, more efficient review experience for others.}       
\fsequote{196051220}{better incentivize what is now thankless volunteer work -- reviewing, replication studies etc. It is not sustainable for people to be doing these things well and frequently without proper incentives. reviewers are}
\fsequote{196061190}{It fails to produce more influential work and new techniques}
\fsequote{196067582}{By comparing to other fields, I think we are too critical to our colleagues research in this sense that we are reluctant to accept papers that replicate previous study, or just provide evidence in the areas that were already studied (sometimes having single studies only) as we believe these are not new results. The same relates to replications - see medical sciences as counter example - they publish single cases as they allow to build body of konowledge in the long run. Also, we are very picky on what improvement is - look at ML field - they have a race on improving things - we expected only novel ideas. I think this makes SE research body of knoweldge very shallow and with little relevancy to the industrial practice...}
\fsequote{196069944}{I would change the review process to heavily weight and reward the replication and extension of prior foundational work, hoping this would establish more reliable results and make SE research cumulative rather than trend-driven.}
\fsequote{196070836}{Despite close ties with industry in some areas, much SE research still struggles to have a tangible impact on industrial practice. The “publish or perish” culture places heavy emphasis on publication counts and conference rankings rather than sustained, high-quality contributions. Although open science is improving, reproducibility is still uneven across the community. The SE community is large and diverse, but that diversity can sometimes lead to fragmentation. The community is still struggling to achieve equitable representation across gender, geography, and economic regions. Many research projects are short-term, driven by grants that last 2–3 years, which limits continuity. }
\fsequote{196071861}{When my paper was nominated as best-paper of a conference. When my PhD students finish their thesis. Also, novel and challenging topics, like ealier energy efficiency and now LLMs engage me with the SE research.}       
\fsequote{196072448}{AI-for-all. Replicability}
\fsequote{196077577}{One issue that does not work well is how some reviewers approach novel or unconventional ideas. In several cases, reviewers seem unwilling to adjust their perspective or engage deeply with the author’s framing. Instead, they hold firmly to their own assumptions or preferred methodologies. This can lead to biased evaluations, where papers proposing new directions or challenging established norms face unnecessary resistance. As a result, the review process can unintentionally discourage innovation and reinforce existing paradigms.}
\fsequote{196098713}{Many (most?) papers do not address key Software Engineering problems. Requirements, IDEs, Visualization, and Debugging are examples of extremely important concerns in SE poorly represented. Often the focus is higher on the solution space (in defense of novelty) as opposed to the problem.}
\fsequote{196144420}{There is maybe too much focus on publishing papers, and moving to the next hot topic, instead of ensuring reproducibility, building platforms to facilitate future research, or revisit topics from new points of view.}
\fsequote{196173071}{One aspect that does not work well in the software engineering research community is the limited reproducibility of many published studies. Complex toolchains, unavailable datasets, and insufficient documentation often make it difficult for other researchers to replicate results. This slows scientific progress, reduces trust in empirical findings, and creates barriers for newcomers trying to build on prior work.}
\fsequote{196181126}{A shared repository for research ideas, datasets, and research materials that lead to better distribution, collaboration, and replication. Authors of such materials should be acknowledged in every research using them.}
\fsequote{196190226}{Academia proposes not so much comprehensive techniques also tool support for software games, evalaution UI/UX design and betting agents because this is new era and perhaps researchers not target this.}
\fsequote{196196304}{1) Better differentiate conferences from journals, with review criteria for the former more unbalanced toward novelty and reproducibility, while for the latter more rewarding indeed novelty AND rigor, soundness, impact and replicability at scale. 2) Foster discussion at conferences, currently limited by i) short presentation time, ii) systematically having non-expert presenters. As for the latter, while it's important to give students the chance to grow, there could be dedicated formats like early-career researchers presentation (besides student forums) and, at the same time, incentivize expert presenters on high-impact tracks, e.g., like with distinguished papers. The desired outcome of both actions, along with stronger focus on novelty in reviews, is to conferences back at places with really open and participated discussions on ideas, without the need of reading hundreds of papers from the proceedings, and to foster cooperation.}
\fsequote{196196304}{The review and publication culture is becoming overly focused on incremental but sound engineering advances, at the expense of novel/disruptive ideas as well as at the expense of replication studies, negative results. Especially conferences like ICSE, FSE, and ASE, as venues of discussion of ideas, should be more open to contributions not yet fully validated in terms of external validity, but with great novelty potential (I would even say not mature but with great potential) – acceptance rates should be a gate for quality of ideas, not a check to exclude studies not yet validated universally (although criteria always state 'look for reasons to accept, rather than reject', we have not achieved that culture yet, as 'good reviews' is often interpreted as 'finding more problems in the paper'). In contrast, journals are the venues where a full mature work should land. I find many cases in which reproducibility is only 'apparent', with researchers even removing files from their Github repos after publication, making their work inaccessible or unusable. I don't have a solution, but this is indeed a barrier to openness.}
\fsequote{196205708}{A community where high-quality, durable artefacts and real-world impact are valued as much as novelty} 
\fsequote{196205708}{Reproducibility is still a challenge, not Fully Welcoming of All Research Branches}
\fsequote{196208899}{Shift toward social aspects, which I do not think most of the software engineering researchers are competent on; too much emphasis on observing actual software engineers rather than on producing new contributions that could help them in their work; cumbersome review processes in conferences, which do not seem to result in significant advancements in the quality of accepted papers}
\fsequote{196208899}{The difficulty of publishing new ideas and contributions as opposed to observational papers (those that study a phenomenon and provide some statistics on it). As a result, younger generations tend to favour the second kind of research rather than the first one or to leave the community and start publishing elsewhere (e.g., distributed systems and cloud computing).}
\fsequote{196247974}{A reviewing practice that does not look for reasons to reject but for reasons to accept. Accepting novelty also comes with replications.}
\fsequote{196247974}{Dissemination of negative results. Replication towards consolidation of ideas and solutions}
\fsequote{196248764}{Lack of reproducibility of some work}
\fsequote{196286246}{Submission inflation (due to publish or perish, probably). It is driving up the review load and, unfortunately, down the review quality. It is also making really hard to stay up to date with relevant breakthrough, which are more and more at risk of getting lost in the noise. The state of reproducibility/replicability in empirical studies is also very sad.}
\fsequote{196289904}{Contamination by AI, especially marginally interesting incremental contributions conservatism is privileged: a dumb but by-the-book paper has more chances to be accepted than one having really new ideas}
\fsequote{196290588}{be more open for paper submissions that may be just about specific case studies, or have negative results, or just replicate previous studies under some different settings.}
\fsequote{196290588}{to keep up with 'novelty' and 'generalizability' requirements for paper submissions; it'd be helpful for the community to be more open for work that may be replication, or a case study, or application of known techniques for a different scenario.} 
\fsequote{196310506}{The artifact evaluation only works on whether the tool can reproduce the experimental data. But as members of the SE community, we all know that the code quality and the robustness of the tools are equally important as the experimental results. Besides, for paper reviewers, the randomly assigned artifact evaluator may be less helpful. If the assigned reviewer is an 'expert' on the topic of the paper under review, they can (but sometimes, might) check the source code, data, and results in detail and provide helps for other reviewers to evaluate the paper; whereas for 'informed outsiders', they can hardly provide helpful suggestions to other reviewers in terms of the quality of the artifact.}
\fsequote{196368894}{Too many papers published lack implementations available on open-source repositories for reproducibility and reuse. Too many papers published evaluate the ideas they put forward on toy case studies rather than scale-realistic case studies coming from genuine industrial needs. Too many software engineering research tools do not practice in their own implementation the engineering principles they purport to support.}
\fsequote{196380886}{Excessive focus on 'novelty' that encourages publishing very low-impact work (e.g., all kinds of 'AI for X' low-hanging fruits), and aversion to work related to engineering/scalability/applicability that makes more real-world impact without necessarily inventing a new wheel.}       
\fsequote{196391274}{A large portion of work in ML4Code, defect prediction, and program analysis continues to overfit small or outdated datasets (such as Defects4J), which limits the real-world generalizability of reported results. At the same time, many SE papers introduce solutions that rarely see adoption in practice, including overly complex static analyses, heavyweight transformation tools, and niche testing frameworks. Across major venues, the incentive structure often favors incremental algorithmic novelty over practical relevance, even though industry impact would benefit far more from robustness, reproducibility, and engineering insights. In addition, conference participation remains challenging for researchers from certain countries due to stringent visa requirements, which can hinder diversity, collaboration, and the overall inclusiveness of the SE community.}
\fsequote{Discord1}{Great point! To me it seems that no - we do not share visions with most  practitioners. We have very different motivations/incentives and it is very hard to create "win-win" scenarios for collaboration. There are a lot of reasons for that. One notable "problematic area"  is that we focus on "novelity" so much that we create many tool prototypes that are unusable in practice - ones thar require painful setup and work mainly on "toy examples". But making a good tool and maintaining it does not really lead to more papers, so why do it? It does "only" lead to great citations on one paper in the best case scenario.}
\fsequote{Discord2}{This might be spicy, but I would posit that academic work in chemical, mechanical, or electrical engineering are generally finding new insights and approaches in chemistry, materials science, etc. My feeling is that most SE research is more like anthropology or demographics: SE literature seems to study "here are properties of groups of engineers, or codebases (MSR)." It feels rare that SE research is about new approaches in software (designs, architecture), or new approaches to efficiency. And that's a shame, because good essays and papers on SE are gold. I cite those things all the time, but a lot of it is coming from informal SE (e.g., anything by Cat Hicks or Nicole Forsgren). Shifting the mass of SE research toward those things can still be academically rigorous and deeply relevant to the industry.}

\end{document}